\documentclass[fleqn,usenatbib]{mnras}

\usepackage{newtxtext,newtxmath}
\usepackage[T1]{fontenc}

\DeclareRobustCommand{\VAN}[3]{#2}
\let\VANthebibliography\thebibliography
\def\thebibliography{\DeclareRobustCommand{\VAN}[3]{##3}\VANthebibliography}
\usepackage{graphicx}
\usepackage{balance}
\usepackage{amsmath}
\usepackage{microtype}
\usepackage[flushleft]{threeparttable}
\usepackage[dvipsnames]{xcolor}

\title[Simulations of the KT\,Eri nova super-remnant]{Hydrodynamic simulations of the KT\,Eridani nova super-remnant}

\author[M. W. Healy-Kalesh et al.]{M. W. Healy-Kalesh,$^{1}$\thanks{E-mail: M.W.HealyKalesh@ljmu.ac.uk (MWH-K)} M. J. Darnley,$^{1}$ M. M. Shara,$^{2}$ K. M. Lanzetta,$^{3}$ J. T. Garland,$^{2}$ and S. Gromoll$^{4}$
\\
$^{1}$Astrophysics Research Institute, Liverpool John Moores University, IC2 Liverpool Science Park, Liverpool, L3 5RF, UK\\
$^{2}$Department of Astrophysics, American Museum of Natural History, Central Park West at 79th Street, New York, NY 10024, USA\\
$^{3}$Department of Physics and Astronomy, Stony Brook University, Stony Brook, NY 11794-3800, USA\\
$^{4}$Amazon Web Services, 410 Terry Ave. N, Seattle, WA 98109, USA\\
}

\date{Accepted 2023 October 10. Received 2023 October 09; in original form 2023 August 19}

\pubyear{2023}

\begin{document}
\label{firstpage}
\pagerange{\pageref{firstpage}--\pageref{lastpage}}
\maketitle

\begin{abstract}
A nova super-remnant (NSR) is an immense structure associated with a nova that forms when frequent and recurrent nova eruptions sweep up surrounding interstellar material (ISM) into a high density and distant shell. The prototypical NSR, measuring over 100 pc across, was discovered in 2014 around the annually erupting nova M\,31N 2008-12a. Hydrodynamical simulations demonstrated that the creation of a dynamic NSR by repeated eruptions transporting large quantities of ISM is not only feasible but that these structures should exist around all novae, whether the white dwarf (WD) is increasing or decreasing in mass. But it is only the recurrent nova (RNe) with the highest WD masses and accretion rates that should host observable NSRs. KT\,Eridani is, potentially, the eleventh RNe recorded in the Galaxy and is also surrounded by a recently unveiled H$\alpha$ shell tens of parsecs across, consistent with a NSR. Through modelling the nova ejecta from KT\,Eri, we demonstrate that such an observable NSR could form in approximately 50,000 years, which fits with the proper motion history of the nova. We compute the expected H$\alpha$ emission from the KT\,Eri NSR and predict that the structure might be accessible to wide-field X-ray facilities.
\end{abstract}


\section{Introduction} \label{sec:Intro}
Classical novae (CNe) are a type of cataclysmic variable that can suddenly increase in luminosity by over five orders of magnitude before fading to quiescence \citep{2012ApJ...756..107S,2012ApJ...758..121S}. CNe arise from accreting binaries \citep{1954PASP...66..230W} in which hydrogen-rich material streams from a less-evolved companion star \citep[typically subdwarf, subgiant, or red giant;][]{2012ApJ...746...61D} onto a white dwarf (WD) via an accretion disk \citep[see, e.g.,][]{1995cvs..book.....W}. As more matter accumulates, the layer at the base of the accreted envelope is subjected to immense temperatures and compression as a result of the WD's intense surface gravity and consequently becomes electron degenerate. This leads to an irreversible cascade of thermonuclear reactions in the form of a thermonuclear runaway (TNR) that rips through the accreted envelope \citep{1972ApJ...176..169S}. The TNR pushes the envelope towards the Fermi temperature whereby electron degeneracy is lifted and the material expands in response to the high temperatures; this is the nova eruption \citep{1976IAUS...73..155S}. Around 10$^{-4} \ \text{M}_\odot$ of material, travelling between a few hundred and a few thousand kilometers per second, is ejected during the eruption \citep{2001IAUS..205..260O} forming an expanding nova shell \citep[see, for example,][]{1990LNP...369..179W,1995MNRAS.276..353S,2014ASPC..490.....W,2022MNRAS.511.1591T}.

The configuration of the nova binary remains unaltered following an eruption. As such, accretion will resume after the re-establishment of the disk \citep{2007MNRAS.379.1557W} and the whole process will repeat. Recurrent novae (RNe) are the resultant phenomenon. Recent theoretical models \citep{2005ApJ...623..398Y,2014ApJ...793..136K} show that all nova systems can accommodate these repeated eruptions and therefore all CNe are inherently recurrent; they simply have inter-eruptions periods that are far longer than the ${\sim}$100 years of modern astronomical data. However, observationally, RNe are defined as exhibiting more than one eruption from the same system; ten such RNe are located within the Galaxy \citep{2010ApJS..187..275S,2021gacv.workE..44D}, with the other twenty-three residing in M\,31 and the LMC \citep{2020AdSpR..66.1147D}. As with CNe, nova shells exist around RNe; a key difference being collisions between consecutive ejecta, which can produce shock-heating and clumping of material as evidenced around the Galactic RN, T\,Pyxidis \citep{1997AJ....114..258S,2013ApJ...768...48T}.

M\,31N 2008-12a (or 12a) is the most extreme RN known \citep[see, e.g.,][and references therein]{2016ApJ...833..149D,2018ApJ...857...68H,2020AdSpR..66.1147D,2021gacv.workE..44D}. Situated in the Andromeda galaxy, 12a has experienced a nova eruption annually since at least 2008 (Healy-Kalesh et al.\ in prep) equating to a mean recurrence period of 0.99 years \citep{2020AdSpR..66.1147D,2021gacv.workE..44D}. These rapid-fire eruptions are produced as a result of the most massive WD in any known nova system \citep[$1.38 \ \text{M}_{\odot}$;][]{2015ApJ...808...52K} accreting material from its companion at an extreme rate \citep[${\sim}10^{-6} \ \text{M}_{\odot} \ \text{yr}^{-1}$;][]{2017ApJ...849...96D}. The exceptional nature of 12a was further showcased through the discovery of a vast surrounding elliptical structure \citep{2015A&A...580A..45D}. With a projected size of $90 \times 134$ parsecs \citep{2019Natur.565..460D} far larger than any single-eruption nova shell \citep[e.g.,][]{2004ApJ...600L..63B,2012ApJ...756..107S,2012ApJ...758..121S} as well as the majority of supernova remnants \citep{2001ApJ...563..816S}, the first nova super-remnant (NSR) had been uncovered \citep{2019Natur.565..460D}.

\citet{2019Natur.565..460D} demonstrated, through hydrodynamical modelling \citep*[utilising the \texttt{MORPHEUS} code;][]{2007ApJ...665..654V}, that an extensive NSR could be created through the continual sweeping up of local interstellar medium by the rapid-fire eruptions of 12a over the system's lifetime. Extensive simulations presented in \citet{2023MNRAS.521.3004H} exploring the influence of several system parameters (including ISM density, accretion rate, WD temperature, and initial WD mass) on NSR growth found that all RNe should be surrounded by vast dynamic NSRs. However, 12a had remained the only nova known to host a NSR.

KT\,Eridani (Nova Eridani 2009) was a bright ($V\sim5.42$), very fast-fading ($t_3\sim13.6$\,d) nova \citep{2022MNRAS.517.3864S} discovered on November 25 2009 \citep{2009IAUC.9098....1Y,2009CBET.2050....1I}. Comprehensive photometric coverage of the pre- and post-maximum light curve with the Solar Mass Ejection Imager \citep[SMEI;][]{JACKSON1997441} instrument on the {\it Coriolos} satellite captured the elusive pre-maximum halt phase of novae evolution \citep{2010ApJ...724..480H,2010ATel.2558....1H}. Though observationally classified as a CN, KT\,Eri is often regarded as the eleventh RN in the Galaxy on account of the system's characteristics, rather than the detection of a second eruption \citep{2014ApJ...788..164P,2022MNRAS.517.3864S}. Specifically it is very fast-fading; has a small outburst amplitude; exhibits spectra containing triple-peaked H$\alpha$ emission with FWHM of 3200--3400 km s$^{-1}$ and high excitation lines (\ion{He}{ii}); has two light curve plateaus and has a WD mass of $1.25\pm0.03\text{M}_{\odot}$ (>1.25$\text{M}_{\odot}$) orbiting a subgiant companion (T$_{\text{eff}} = 6200$\,K) with an orbital period of $P = 2.61595$\,d \citep{2014ApJ...788..164P,2022MNRAS.517.3864S}. Furthermore, \citet{2022MNRAS.517.3864S} use their determinations of the WD mass and accretion rate ($3.5 \times 10^{-7} \ \text{M}_{\odot} \ \text{yr}^{-1}$) to derive a recurrence period of 40--50 years. 

Narrowband imaging of the surroundings of KT\,Eri,  obtained with the Condor Array Telescope \citep{2023PASP..135a5002L}, reveal a vast structure, ${\sim}50$ parsecs in diameter, coincident with KT\,Eri (Shara et al.\ 2023), akin to 12a's NSR \citep{2019Natur.565..460D}.

In this Paper, we model nova ejecta from a fixed mass WD in an attempt to replicate the NSR observed surrounding KT\,Eri. In Section~\ref{sec:simulations} we present our hydrodynamic simulations before presenting the results in Section~\ref{sec:Results}. We then discuss consistencies between our modelling and the observed NSR in Section~\ref{sec:Discussion} before concluding our paper in Section~\ref{sec:Conclusions}.

\begin{table}
\caption{Parameters for each simulation. Columns record the simulation number, recurrence period, ejecta velocity, number of eruptions to grow the nova super-remnant to 25 parsecs (the angular radius of the observed NSR around KT\,Eri - see Shara et al.\ 2023), the elapsed (cumulative) time of the simulation, and the total kinetic energy released. Runs 1$^\star$ -- 5$^\star$ have the same characteristics as Runs 1 -- 5 but the single component ejecta has been replaced by multicomponent ejecta as described in Section~\ref{sec:Multicomponent ejecta}. Run 4$^\dag$ has the same characteristics as Run 4 however the recurrence period differs from that given as set by a Gaussian distribution (see Section~\ref{sec:Gaussian distribution recurrence periods} for details).}
\label{Runs}
\begin{center}
\begin{tabular}{cccccc}
\hline
Run \# & $\text{P}_{\text{rec}}$ & $v_{\text{ej}}$ & Number of & Elapsed time & Total KE\\
& (yrs) & (km/s) & eruptions & (years) & (erg) \\
\hline
\phantom{0}1 & \phantom{00}5 & 6,000 & 10,113 & $5.1 \times 10^{4}$ & $6.5 \times 10^{48}$ \\
\phantom{0}2 & \phantom{0}10 & 6,000 & \phantom{0}5,076 & $5.1 \times 10^{4}$ & $6.5 \times 10^{48}$ \\
\phantom{0}3 & \phantom{0}20 & 6,000 & \phantom{0}2,543 & $5.1 \times 10^{4}$ & $6.5 \times 10^{48}$ \\
\phantom{0}4 & \phantom{0}50 & 6,000 & \phantom{0}1,019 & $5.1 \times 10^{4}$ & $6.5 \times 10^{48}$ \\
\phantom{0}5 & 100 & 6,000 & \phantom{0,0}510 & $5.1 \times 10^{4}$ & $6.5 \times 10^{48}$ \\
\phantom{0}6 & \phantom{00}5 & 5,000 & 11,435 & $5.8 \times 10^{4}$ & $5.1 \times 10^{48}$ \\
\phantom{0}7 & \phantom{0}10 & 5,000 & \phantom{0}5,739 & $5.8 \times 10^{4}$ & $5.1 \times 10^{48}$ \\
\phantom{0}8 & \phantom{0}20 & 5,000 & \phantom{0}2,875 & $5.8 \times 10^{4}$ & $5.1 \times 10^{48}$ \\
\phantom{0}9 & \phantom{0}50 & 5,000 & \phantom{0}1,152 & $5.8 \times 10^{4}$ & $5.1 \times 10^{48}$ \\
10 & 100 & 5,000 & \phantom{0,0}576 & $5.8 \times 10^{4}$ & $5.1 \times 10^{48}$ \\
11 & \phantom{00}5 & 4,000 & 13,262 & $6.7 \times 10^{4}$ & $3.8 \times 10^{48}$ \\
12 & \phantom{0}10 & 4,000 & \phantom{0}6,656 & $6.7 \times 10^{4}$ & $3.8 \times 10^{48}$ \\
13 & \phantom{0}20 & 4,000 & \phantom{0}3,334 & $6.7 \times 10^{4}$ & $3.8 \times 10^{48}$ \\
14 & \phantom{0}50 & 4,000 & \phantom{0}1,336 & $6.7 \times 10^{4}$ & $3.8 \times 10^{48}$ \\
15 & 100 & 4,000 & \phantom{0,0}668 & $6.7 \times 10^{4}$ & $3.8 \times 10^{48}$ \\
\hline
\phantom{0}1$^\star$ & \phantom{00}5 & variable & 10,113 & $5.1 \times 10^{4}$ & $6.5 \times 10^{48}$ \\
\phantom{0}2$^\star$ & \phantom{0}10 & variable & \phantom{0}5,076 & $5.1 \times 10^{4}$ & $6.5 \times 10^{48}$ \\
\phantom{0}3$^\star$ & \phantom{0}20 & variable & \phantom{0}2,543 & $5.1 \times 10^{4}$ & $6.5 \times 10^{48}$ \\
\phantom{0}4$^\star$ & \phantom{0}50 & variable & \phantom{0}1,019 & $5.1 \times 10^{4}$ & $6.5 \times 10^{48}$ \\
\phantom{0}5$^\star$ & 100 & variable & \phantom{0,0}510 & $5.1 \times 10^{4}$ & $6.5 \times 10^{48}$ \\
\hline
\phantom{0}4$^\dag$ & \phantom{0}50 & 6,000 & \phantom{0}1,033 & $5.2 \times 10^{4}$ & $6.6 \times 10^{48}$ \\
\hline
\end{tabular}
\end{center}
\end{table}

\section{Simulations}\label{sec:simulations}
As with previous work simulating NSRs \citep{2019Natur.565..460D,2023MNRAS.521.3004H}, we utilised \texttt{MORPHEUS} \citep{2007ApJ...665..654V} to model the evolution of the NSR shell associated with KT\,Eri. \texttt{MORPHEUS} integrates three codes, namely the one-dimensional Asphere \citep{2007ApJ...665..654V}, two-dimensional Novarot \citep{1997MNRAS.284..137L} and the three-dimensional CubeMPI \citep{2006MNRAS.366..387W}, developed by the Manchester University--LJMU Nova groups to construct an MPI-OpenMP Eulerian second-order Godunov simulation code.

In \citet{2023MNRAS.521.3004H}, the authors followed the growth of NSRs created by nova ejecta from evolving (mass growth or reduction) WDs. Those simulations were concerned with the extreme evolution of NSRs as the WD approached the Chandrasekhar mass. For KT\,Eri, the much longer recurrence periods are close to those during early `steady state' NSR evolution. Therefore, we evolve a NSR from a fixed mass WD (mass accumulation efficiency = 0) in a similar manner to the simulations of the 12a NSR by \citet{2019Natur.565..460D}.

We adopted the same mass loss rate of $2 \times 10^{-8} \ \text{M}_{\odot} \ \text{yr}^{-1}$ and wind velocity of 20 km s$^{-1}$ from the companion as used in the previous studies \citep[see][for details]{2019Natur.565..460D,2023MNRAS.521.3004H}. While KT\,Eri has a companion with an effective temperature that is consistent with a subgiant \citep{2022MNRAS.517.3864S}, we use this mass loss rate from the companion between nova eruptions predominantly for its help with computational time. The eruptions from the central nova are then represented by a instantaneous increase in ejecta velocity and mass-loss rate tuned to match the eruption of KT\,Eri (as described in Section~\ref{sec:Intro}), punctuating the wind mass-loss at intervals given by a set recurrence period. As in previous studies, we assume one-dimensional spherical symmetry for these simulations (largely for computational reasons). For our simulations, we have chosen a resolution of 1000 au/cell as this is able to showcase the gross structure of the NSR during its full evolution (see Section~\ref{sec:Reference simulation dynamics} and Figure~\ref{fig:change resolution ref sim} for a comparison between different spatial resolutions), while allowing for feasible computational times. Furthermore, the resolution we have chosen is so much greater than the orbital separation of the WD and the companion in the KT\,Eri system \citep[10.4\,R$_{\odot}$;][]{2022MNRAS.517.3864S} that we can ignore any ejecta-donor/accretion disk interaction.

\subsection{Radiative cooling}\label{sec:Radiative cooling}
The role of radiative cooling in the formation of the 12a NSR was first explored in \citet{2019Natur.565..460D}. They found that cooling had little influence on NSR evolution as the annual eruptions were highly energetic throughout the full evolution and, did not allow sufficient time for the NSR to cool. In contrast, \citet{2023MNRAS.521.3004H} found that radiative cooling has a large influence on formation of NSRs (including its radial size and shell structure) created by eruptions from an evolving WD; this is a result of the long inter-eruption periods and low energy eruptions during the early `steady state' growth permitting sufficient time for effective cooling. As such, we incorporate the same radiative cooling in this work as employed in previous work \citep{2007ApJ...665..654V,2019Natur.565..460D,2023MNRAS.521.3004H} whereby the \citet{1976ApJ...204..290R} cooling curve within \texttt{Morpheus} is adopted.

\subsection{Local ISM density}\label{sec:Local ISM density}
The density of the surrounding ISM plays a pivotal role in shaping the growing NSR \citep{2023MNRAS.521.3004H}. As such, we need to estimate the ISM density of the immediate surroundings of KT\,Eri. KT\,Eri is situated at a {\it Gaia} distance of $5110^{+920}_{-430}$ pc \citep{2022MNRAS.517.3864S} and lies ${\sim}3$ kpc below the Galactic plane (Shara et al.\ 2023). The scale height of gas in the peripheries of Milky Way-like galaxies is ${\sim}800$ pc \citep{2023MNRAS.518L..63G} therefore KT\,Eri lies approximately $4 \times$ this scale height from the Galactic plane. This indicates that the gas density in the region where KT\,Eri is located is ${\sim}2$ per cent of the gas in the plane of the Galaxy in the neighbourhood of KT\,Eri. Direct measurements by the Voyager 1 spacecraft \citep{2023ApJ...951...71K} find the local ISM electron density (and therefore likely hydrogen) in the Galactic plane to be ${\sim}0.1\ \text{cm}^{-3}$. Therefore, we predict the ISM density around KT\,Eri to be $2 \times 10^{-3}$ H atom per cubic centimetre. This ISM density is utilised for all models in this study. For consistency with models in \citet{2023MNRAS.521.3004H}, we refer to this ISM density by the number density $n = 2 \times 10^{-3} \ \text{cm}^{-3}$ (dropping the units) throughout.

\subsection{Bulk ejecta simulations}\label{sec:Bulk ejecta simulations}
The first set of simulations (Runs 1 -- 15) follow the growth of a NSR from the central nova ejecting a single bulk of material with every eruption. While a simplistic interpretation of a nova eruption, \citet{2023MNRAS.521.3004H} showed that the gross structure of a NSR is unaffected by the structure of each ejecta (see Section~\ref{sec:Multicomponent ejecta} for the possible impact of a complex-multi-component ejecta on NSR substructure).

We assume a static WD for our study as (i) suitable eruption models for KT\,Eri do not yet exist, (ii) it is relatively poorly studied (compared to other RNe) and (iii) its long recurrence period suggests KT\,Eri may be a relatively young system compared to other RNe \citep[e.g., RS\,Ophiuchi, U\,Scorpii and M\,31N 2008-12a; see e.g.,][]{2021gacv.workE..44D}.

For the static WD scenario, the total mass ejected during each nova eruption can be found by taking the mass accretion rate of KT\,Eri \citep[$3.57 \times 10^{-7} \ \text{M}_{\odot} \ \text{yr}^{-1}$;][]{2022MNRAS.517.3864S} over the course of an inter-eruption period. The recurrence period of KT\,Eri is predicted to be 40 -- 50 years by \citet{2022MNRAS.517.3864S}, however, with only one observed eruption of KT\,Eri, other recurrence periods are viable. As such, we chose to sample a range of recurrence periods: 5, 10, 20, 50 and 100 years. 

For the velocity of the nova ejecta, \citet{2013IAUS..281..119A} report that during its nebular phase, KT\,Eri exhibited [\ion{O}{iii}] $\lambda4959$ and $\lambda5007$ lines composed of multiple components with velocities of $-2000 \ \text{km} \ \text{s}^{-1}$, $-1000 \ \text{km} \ \text{s}^{-1}$, $+700 \ \text{km} \ \text{s}^{-1}$ and $+1800 \ \text{km} \ \text{s}^{-1}$. Additionally, \citet{2009IAUC.9098....1Y} report broad Balmer lines with the H$\alpha$ emission line FWHM being approximately $3200 - 3400 \ \text{km} \ \text{s}^{-1}$ alongside broad emission lines in the $0.9\, \mu$m to $2.5\, \mu$m regime with velocities of $4000 \ \text{km} \ \text{s}^{-1}$. KT\,Eri is classified as a fast He/N nova, therefore as well as an ejecta velocity of $4000 \ \text{km} \ \text{s}^{-1}$ (in line with the velocities derived from spectra), we also chose to model eruptions with feasible ejecta velocities of $5000 \ \text{km} \ \text{s}^{-1}$ and $6000 \ \text{km} \ \text{s}^{-1}$.

The ejecta mass, recurrence period and ejecta velocities outlined were adopted to construct the characteristics of the nova ejecta in each of our models. The combinations of parameters for Runs 1 -- 15 are detailed in Table~\ref{Runs}.
\begin{figure*}
\includegraphics[width=\textwidth]{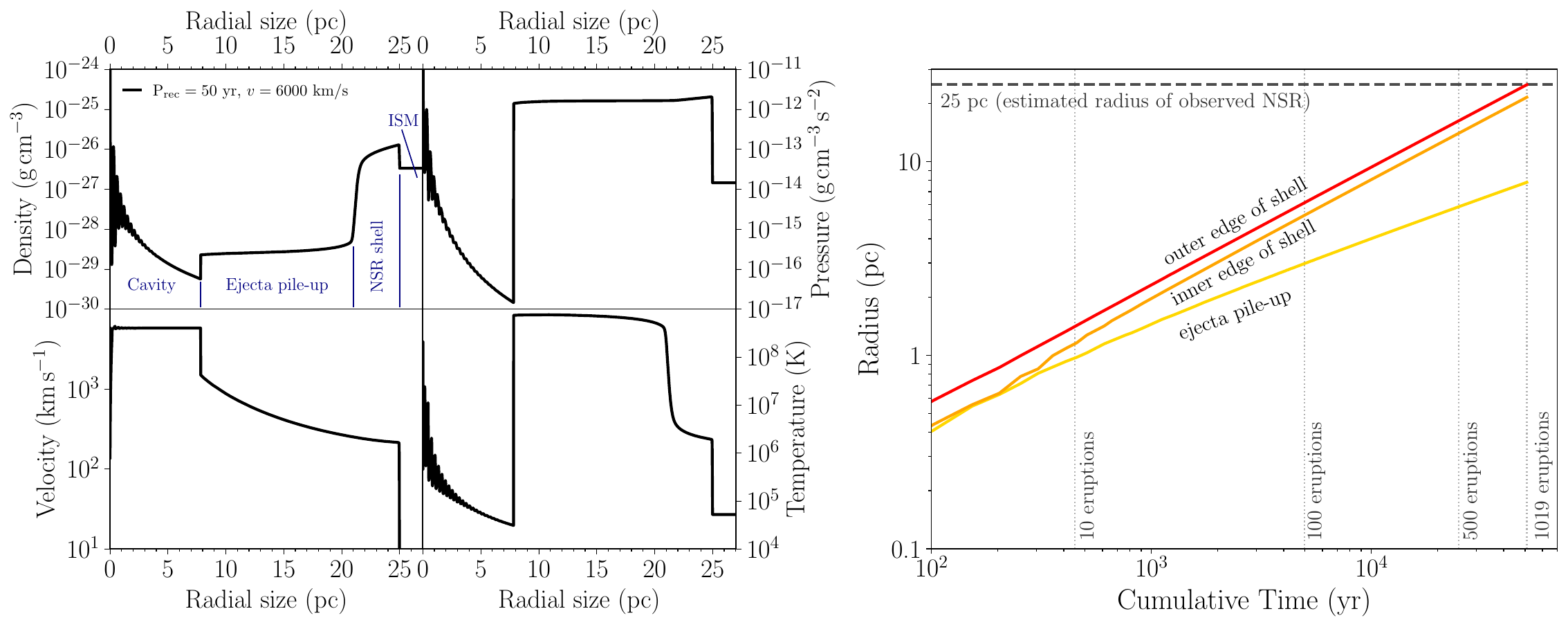}
\caption{Left: The dynamics of the reference simulation (Run 4) NSR with $3.57 \times 10^{-5} \ \text{M}_{\odot} \ \text{yr}^{-1}$ and $n = 2 \times 10^{-3}$ after ${\sim}51$ kyr (1019 eruptions) with 1000 au resolution. The regions of interest we refer to in the text are labelled. Right: The evolution of the inner edge of the ejecta pile-up region and the inner and outer edges of the NSR shell with respect to cumulative time. The estimated radius of the observed shell around KT\,Eri is indicated. \label{fig:ref sim evolution}}
\end{figure*}

\subsection{Multicomponent ejecta}\label{sec:Multicomponent ejecta}
Nova eruptions are comprised of multiple components of ejecta: material being ejected initially at slower velocities followed by faster ejecta at later times \citep{1994MNRAS.271..155O}. This leads to intra-ejecta shocks that heavily contribute to the total optical luminosity of the nova event \citep{2020NatAs...4..776A,2020ApJ...905...62A,2022MNRAS.514.6183M}.

Accordingly, the second set of simulations (Runs 1$^{\star}$ -- 5$^{\star}$) mimic the broad characteristics of Runs 1 -- 5, however the single bulk ejecta detailed in Section~\ref{sec:Bulk ejecta simulations} are replaced by a multicomponent ejecta. Specifically, each eruption is broken down into twelve distinct components, with linearly decreasing ejecta velocities from 6000 km\,s$^{-1}$ to 0 km\,s$^{-1}$ over the same timescale as the 3 mag decline time ($t_3 \sim 13.6$d).

The total kinetic energy contained within the nova eruptions influences the growth of the NSR and its ultimate size \citep{HealyPhD,2023MNRAS.521.3004H}. Therefore, we scaled the total ejected mass from the combined components of an eruption such that its total KE matched the total KE of the eruption with bulk ejecta (as in Section~\ref{sec:Bulk ejecta simulations}). The details of these runs are also presented in Table~\ref{Runs}.

\section{Results}\label{sec:Results}
\subsection{Reference simulation: Run 4}\label{sec:Reference simulation}
\subsubsection{Dynamics}\label{sec:Reference simulation dynamics}
We have selected Run 4 as our reference simulation given that the recurrence period of 50 years is closest to the predicted value as given in \citet{2022MNRAS.517.3864S}. Each eruption from this system ejects $1.785 \times 10^{-5} \ \text{M}_{\odot}$ at a velocity of $6000 \ \text{km} \ \text{s}^{-1}$ into a prepopulated ISM with a density of $n = 2 \times 10^{-3}$ ($3.34 \times 10^{-27} \ \text{g} \ \text{cm}^{-3}$). We find that it would take this nova system $5.1 \times 10^{4}$ years over 1019 eruptions to grow a NSR to 25 pc -- the radial size of the shell surrounding KT\,Eri (Shara et al.\ 2023). 

In the left-hand plot of Figure~\ref{fig:ref sim evolution}, we show the dynamics of the NSR after the full $5.1 \times 10^{4}$ years including its radial density, pressure, temperature and velocity distribution. We also indicate the locations of the cavity, ejecta pile-up region and the NSR shell (confined by its inner and outer edge) that we refer to throughout the paper. The radial growth curve of the outer and inner of the shell and the ejecta pile-up boundary in provided in the right-hand plot of Figure~\ref{fig:ref sim evolution}.

We can see in the top left-hand panel of the left plot of Figure~\ref{fig:ref sim evolution} that, alongside the outer edge of the NSR shell extending to a radius of 25 pc, the inner edge of the shell reaches out to ${\sim}21.6$ pc; this corresponds to a shell thickness of ${\sim}13.7$ per cent. As evident in the right-hand plot of Figure~\ref{fig:ref sim evolution}, the thickness of the NSR shell remains approximately constant throughout its full evolution: e.g., $14.2$ per cent and $13.6$ per cent after 100 eruptions (${\sim}4,950$ yr) and 500 eruptions (${\sim}24,970$ yr), respectively. A constant NSR shell thickness is reminiscent of the NSR grown around the 12a system in \citet{2019Natur.565..460D} and during the early `steady state' evolution from \citep{2023MNRAS.521.3004H} .

Near to the central nova, the density is high as individual eruptions eject mass into the established structure (see top left-hand panel of the left plot in Figure~\ref{fig:ref sim evolution}). Yet beyond this, the density within the cavity drops to $n \simeq 3.4 \times 10^{-6}$ (${\sim}600\times$ less dense than the surrounding ISM density) at the ejecta pile-up boundary due to the high velocity eruptions traversing the region with very little resistance. With such low densities in the cavity, the newly ejected material experiences close to free-expansion and therefore maintain velocities of ${\sim}5800 \ \text{km} \ \text{s}^{-1}$ out to ${\sim}7.8$ parsec, as illustrated in the bottom left-hand panel of the left plot in Figure~\ref{fig:ref sim evolution}. The pressure (see top right-hand panel) and temperatures (see bottom right-hand panel) also drop precipitously across the cavity in response to the sparse levels of material here.

After crossing the low-density cavity, the ejecta collide with the ejecta pile-up region at ${\sim}7.8$ parsec, resulting in velocities dropping considerably to ${\sim}1800 \ \text{km} \ \text{s}^{-1}$. As shown in the left-hand plot of Figure~\ref{fig:ref sim evolution}, while the density increases approximately four-fold at the border, the pressure and temperatures increase substantially to $1.2 \times 10^{-12} \ \text{g} \ \text{cm}^{-3} \ \text{s}^{-2}$ and $7.1 \times 10^8$ K (both over four orders of magnitude), respectively, as collisions between incoming ejecta with the established pile-up region violently shock-heats the material.

Away from the extreme interface of the cavity and pile-up region, the density of material within the ejecta pile-up region remains relatively constant (${\sim}2 - 5 \times 10^{-29} \ \text{g} \ \text{cm}^{-3}$) out to the inner edge of the NSR shell, as does the pressure ($1.2 - 1.6 \times 10^{-12} \ \text{g} \ \text{cm}^{-3} \ \text{s}^{-2}$) and temperature ($4 - 7 \times 10^8$ K), though the temperature of the material drops appreciably ${\sim}5$ pc prior to the shell. On the other hand, due to the ejecta interacting with both material from preceding eruptions and reverse shock fronts, the velocity of the material within the pile-up region drops continuously from ${\sim}1800 \ \text{km} \ \text{s}^{-1}$ down to ${\sim}250 \ \text{km} \ \text{s}^{-1}$ within the NSR shell.

The NSR shell, as indicated in the top left-hand panel of the left plot in Figure~\ref{fig:ref sim evolution}, is a high density (${\sim}1.3 \times 10^{-26} \ \text{g} \ \text{cm}^{-3}$ at the outer edge) region comprised almost exclusively of ISM material swept up by the frequent highly energetic nova eruptions. While only ${\sim}4 \times$ more dense than the surrounding ISM, the density of the shell is up to ${\sim}650 \times$ and ${\sim}2300 \times$ greater than the pile-up region and cavity it encompasses, respectively. In the vicinity closest to the inner edge of the shell, the temperature drops abruptly from ${\sim}2.3 \times 10^8$\,K to ${\sim}5.9 \times 10^6$\,K (at the inner edge of the shell) in the space of approximately 0.5 pc. The temperature gradient between the inner and outer edge (${\sim}1.9 \times 10^6$ K) stems from the material within the shell being more and more shielded from the high energy ejecta impacting the inner edge, and therefore is able to cool. There is a small increase in pressure near to the outer edge (${\sim}2 \times 10^{-12} \ \text{g} \ \text{cm}^{-3} \ \text{s}^{-2}$) with the velocity remaining approximately ${\sim}250 \ \text{km} \ \text{s}^{-1}$ across the whole shell.

\begin{figure}
\includegraphics[width=\columnwidth]{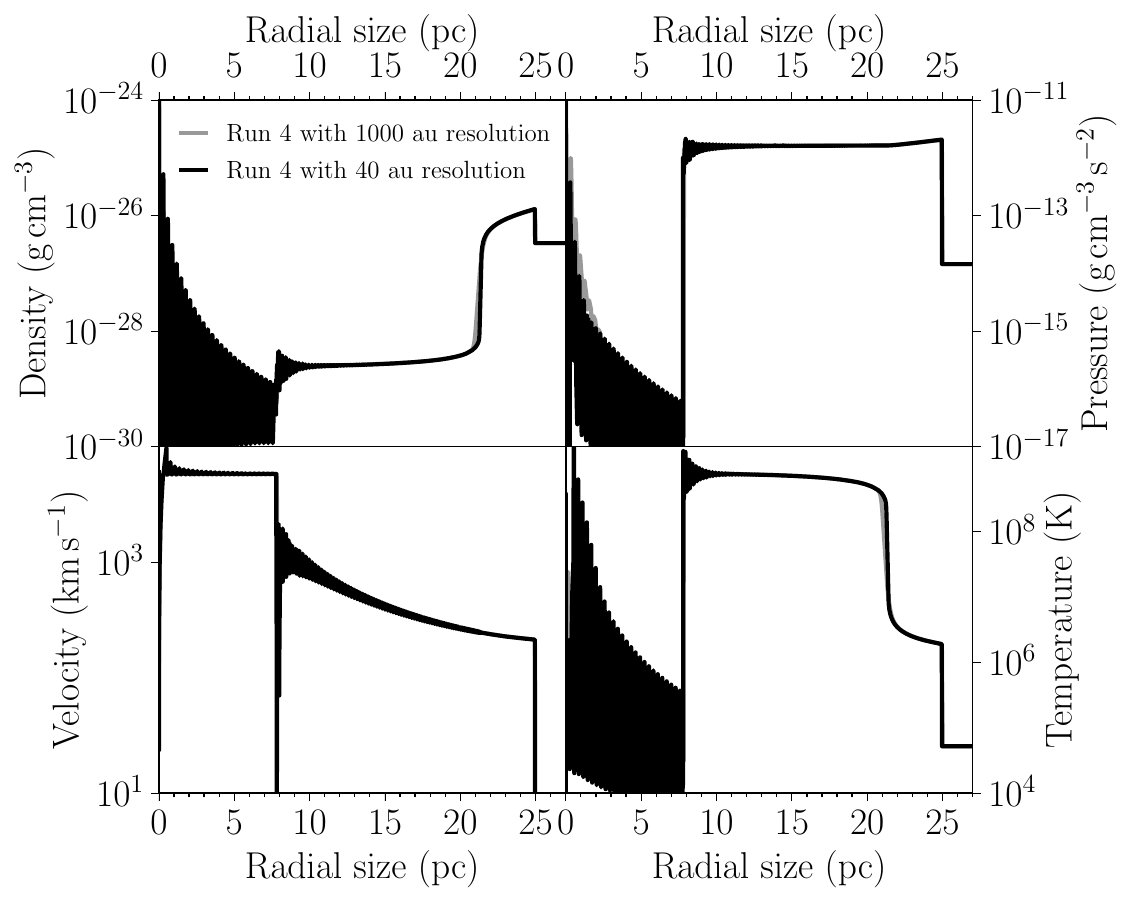}
\caption{Comparing the dynamics of Run 4 (resolution of 1000 au/cell) and Run 4 with a resolution of 40 au/cell. \label{fig:change resolution ref sim}}
\end{figure}

We explore the role of spatial resolution in this study by re-simulating Run 4 with a resolution of 40 au/cell; in Figure~\ref{fig:change resolution ref sim} we compare this higher resolution simulation with Run 4 (1000 au/cell). The spatial resolution chosen for this study does not have an impact on the large-scale structure of the NSR. The most significant difference seen in Figure~\ref{fig:change resolution ref sim} is the expected greater number of resolved individual ejecta across the cavity and pile-up boundary.

\subsubsection{X-ray luminosity}\label{sec:Reference simulation X-ray luminosity}\label{Reference simulation X-ray luminosity}
\begin{figure}
\includegraphics[width=\columnwidth]{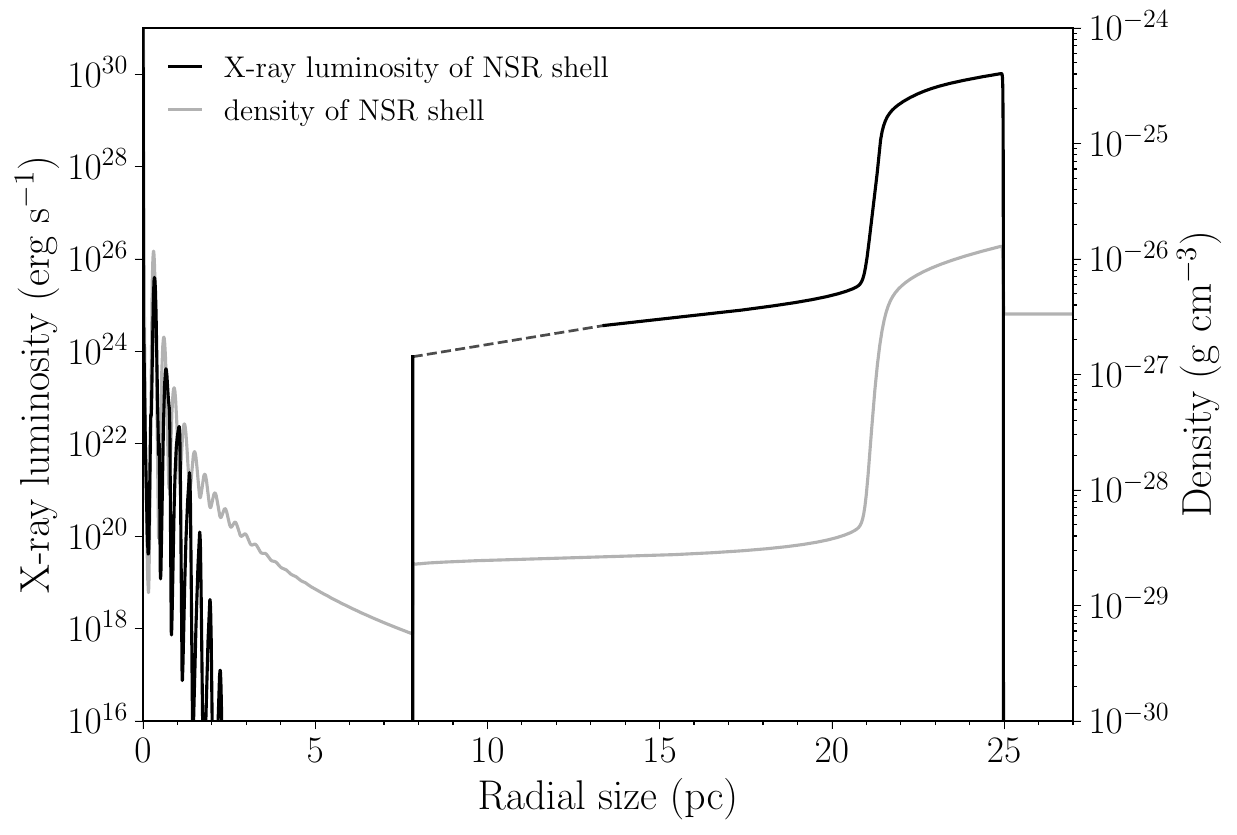}
\caption{Synthetic X-ray luminosity (without absorption) of the NSR grown in Run 4 (black), along with the density distribution of the NSR (grey). The dashed line indicates an interpolation as described in the text. \label{fig:ref sim X-ray}}
\end{figure}
To predict the X-ray luminosity of the NSR grown in Run 4, we adopted the same method as used previous work \citep{Vaytet_Thesis,2011ApJ...740....5V,2019Natur.565..460D,2023MNRAS.521.3004H}. This involves computing the emission measure contribution from each spherical shell of the NSR in Run 4 before binning these contributions into logarithmically divided temperature bins. We then pass these temperature-binned emission measures into \texttt{XSPEC} while employing the \texttt{APEC} \citep{2001ApJ...556L..91S} model (without absorption) to obtain the X-ray luminosity. The radial profile of the NSR X-ray luminosity is shown in Figure~\ref{fig:ref sim X-ray}. We have interpolated the likely X-ray luminosity for the region between ${\sim}7.8$ pc and ${\sim}13.4$ pc (represented as a dashed line in Figure~\ref{fig:ref sim X-ray}) due to the temperatures in this region being too high ($> 64$\,keV) for the \texttt{APEC} model to predict the X-ray emission.

We can see in Figure~\ref{fig:ref sim X-ray} that the X-ray luminosity of the NSR from Run 4 is relatively high (between $\sim10^{20-25} \ \text{erg} \ \text{s}^{-1}$) at the centre as a  result of the highly energetic nova eruptions. Beyond ${\sim}2$ pc, the X-ray luminosity drops away dramatically as the ejecta travel unimpeded through the low density cavity. Once the ejecta crashes into the high density pile-up region border (at ${\sim}7.8$ pc), significant shock-heating occurs leading to a huge increase in X-ray emission ($10^{24} \ \text{erg} \ \text{s}^{-1}$). Shock-heating across the whole ejecta pile-up region maintains the high X-ray luminosity up to the inner edge of the NSR shell. Here, the density increases by over two orders of magnitude -- this triggers another substantial jump in X-ray emission, reaching a peak of $\text{L}_{\text{X-ray}} \simeq 10^{30} \ \text{erg} \ \text{s}^{-1}$ at the edge of the NSR shell. 

The total X-ray luminosity from the whole NSR is $\text{L}_{\text{X-ray}}=6.5 \times 10^{32} \ \text{erg} \ \text{s}^{-1}$, and so is much brighter than the predicted X-ray luminosities of the NSR surrounding 12a: $3 \times 10^{29} \ \text{erg} \ \text{s}^{-1}$ \citep{2019Natur.565..460D} and $1 \times 10^{31} \ \text{erg} \ \text{s}^{-1}$ \citep{2023MNRAS.521.3004H}. Even still, $\text{L}_{\text{X-ray}}$ for Run 4 is over four orders of magnitude fainter than the typical X-ray luminosities of novae during their super-soft source phase \citep[see, e.g.,][]{2010A&A...523A..89H,2011A&A...533A..52H}, however, does exceed the X-ray luminosity of some quiescent novae  \citep[see, e.g., RS\,Oph;][]{2022MNRAS.514.1557P} therefore allowing for potential detection.
\subsubsection{H$\alpha$ surface brightness}\label{sec:Reference simulation Halpha luminosity}
\begin{figure}
\includegraphics[width=\columnwidth]{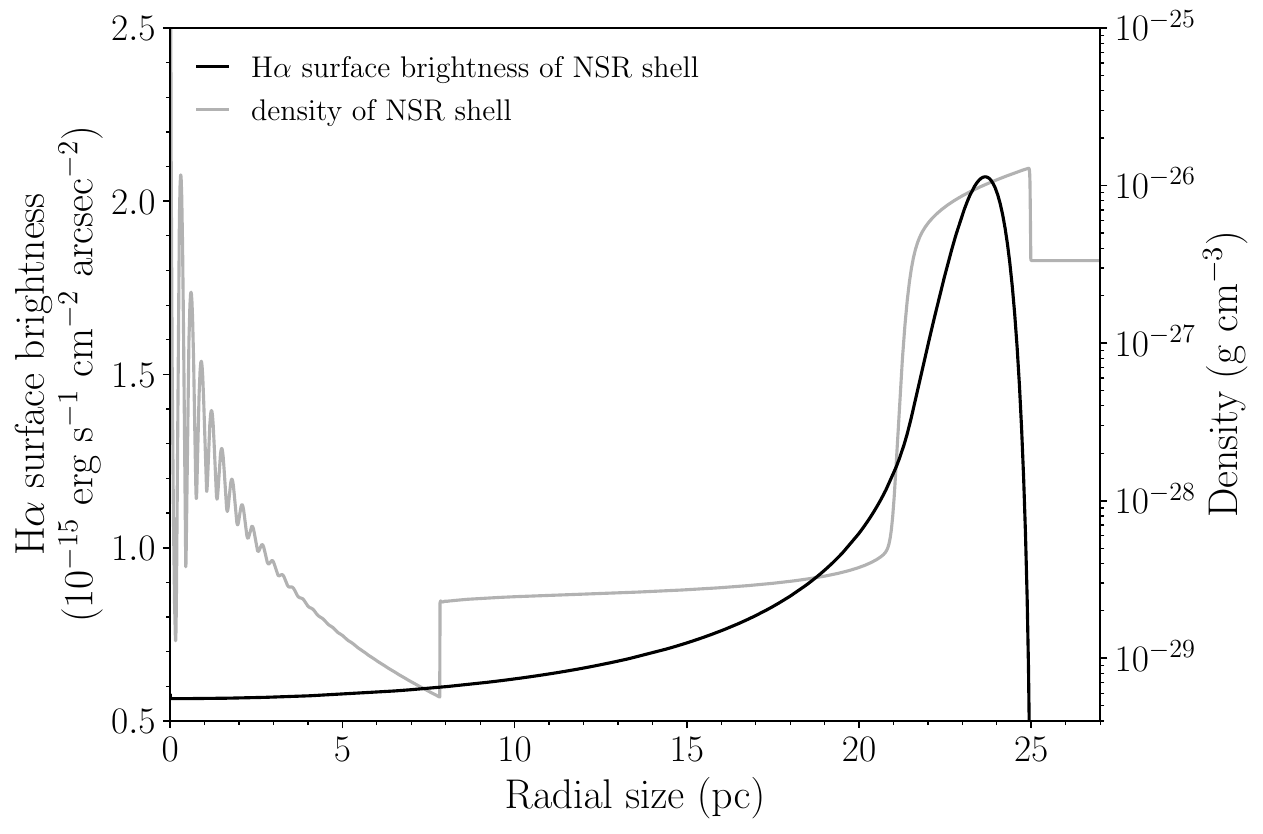}
\caption{As in Figure~\ref{fig:ref sim X-ray} but with the simulated H$\alpha$ surface brightness. \label{fig:ref sim Ha}}
\end{figure}
As with the NSR around 12a \citep{2019Natur.565..460D}, the shell surrounding KT\,Eri is visible through its prominent H$\alpha$ emission (Shara et al.\ 2023). We compute the H$\alpha$ surface brightness of Run 4 by firstly following the method described in \citet{2023MNRAS.521.3004H}. Here, we place the simulated NSR from Run 4 at the distance of KT\,Eri \citep[${\sim}5110$ pc;][]{2022MNRAS.517.3864S} and applied the H$\alpha$ extinction ($A_{\text{H}\alpha} = 0.208$) toward the nova. The total predicted H$\alpha$ luminosity of the NSR from this radial distribution for Run 4 is $\text{L}_{\text{H}\alpha} \simeq 1.1 \times 10^{33} \ \text{erg} \ \text{s}^{-1}$ and the vast majority of this emission emanates from the NSR shell.

The simulated H$\alpha$ surface brightness distribution, assuming a spherical geometry, is shown in Figure~\ref{fig:ref sim Ha}. At all radii, the dominant contribution to the surface brightness is the projected NSR shell. The surface brightness close to the nova is ${\sim}6 \times 10^{-16} \ \text{erg} \ \text{s}^{-1} \ \text{cm}^{-2} \ \text{arcsec}^{-2}$ and gradually increases with increasing radius, reaching ${\sim}1.3 \times 10^{-15} \ \text{erg} \ \text{s}^{-1} \ \text{cm}^{-2} \ \text{arcsec}^{-2}$ at the NSR shell inner edge and then peaks at ${\sim}2.1 \times 10^{-15} \ \text{erg} \ \text{s}^{-1} \ \text{cm}^{-2} \ \text{arcsec}^{-2}$ within the NSR shell at ${\sim}23.7$ pc. Any H$\alpha$ contribution from ISM beyond, or in front or behind, the shell is not included. We note that the predicted H$\alpha$ surface brightness of the shell is ${\sim}20-50$ times larger than that reported by Shara et al.\ (2023). This discrepancy may result from the cooling package employed within our simulations (see Section~\ref{sec:Radiative cooling}) whereby the cooling rates are potentially overestimated for densities $\sim 1 \ \text{cm}^{-3}$ and not scaled for the lower ISM densities. A number of other factors could also have affected this estimate, including the ISM density, the assumption that the ISM is purely hydrogen, the assumed eruption history, but, most importantly, the assumption of spherical symmetry of both the ejecta and the ISM.
\begin{figure*}
\centering
\includegraphics[width=0.49\textwidth]{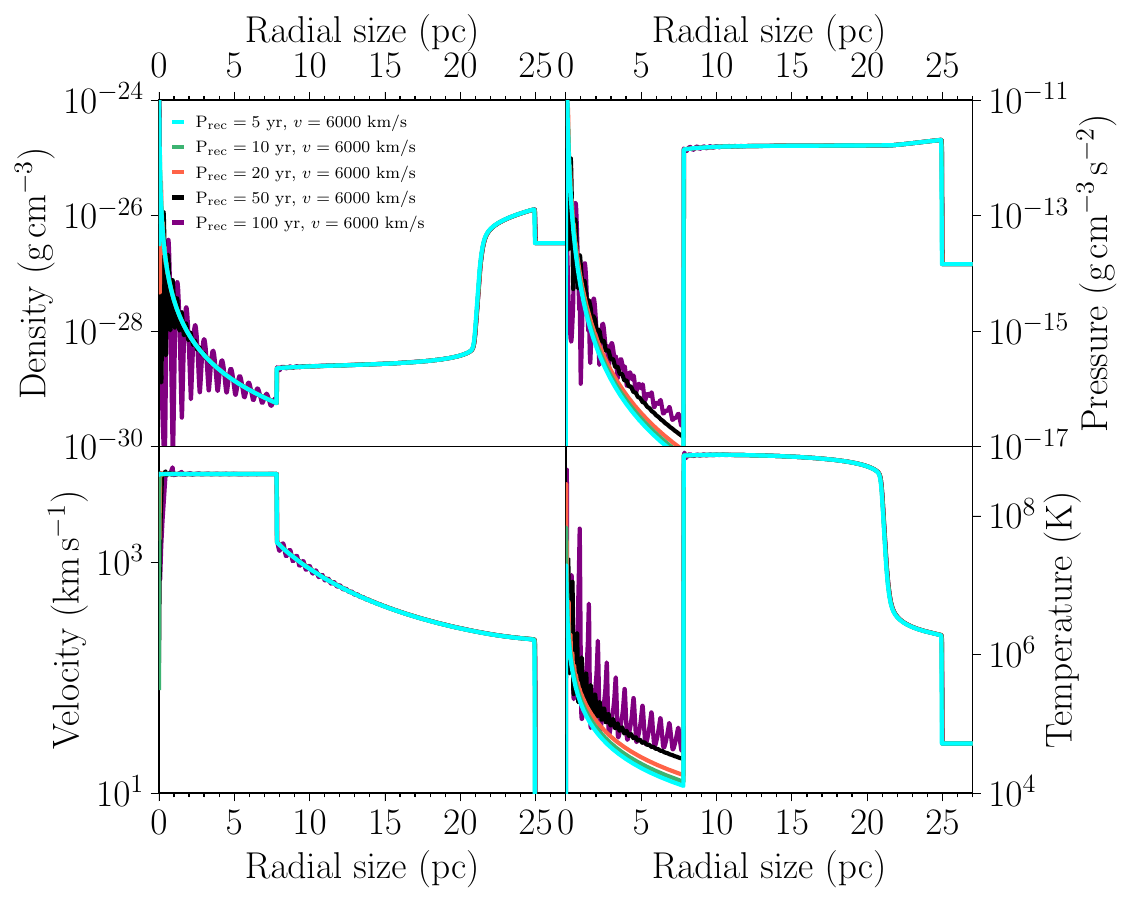} \quad
\centering
\includegraphics[width=0.49\textwidth]{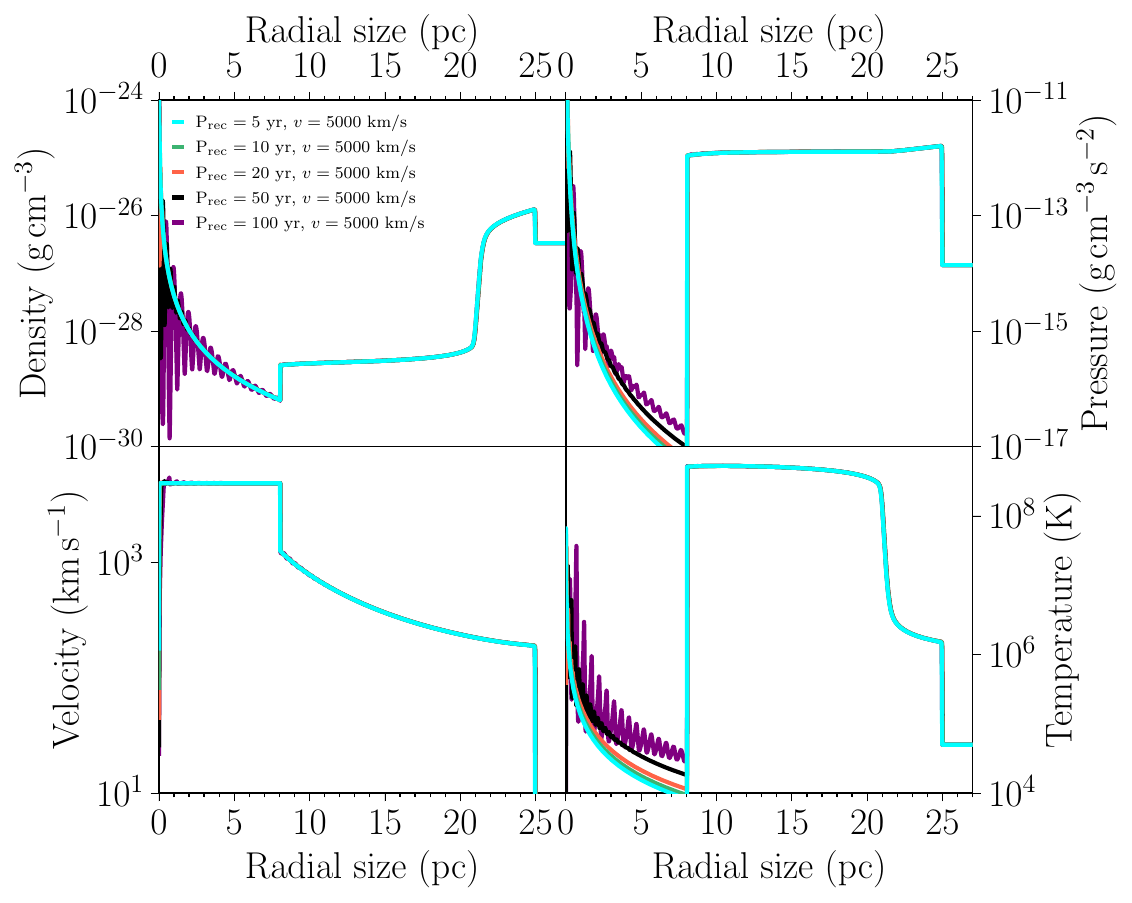} \\
\centering
\includegraphics[width=0.49\textwidth]{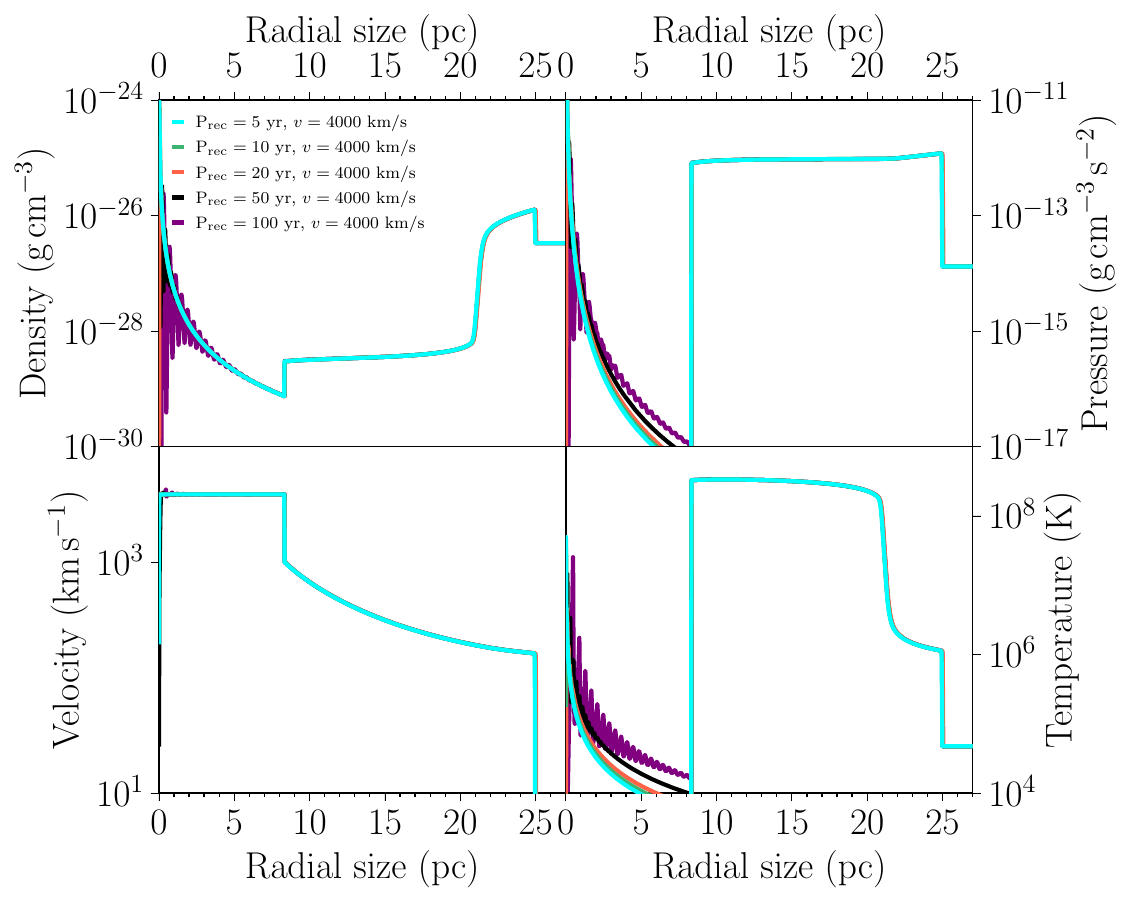} \quad
\centering
\includegraphics[width=0.49\textwidth]{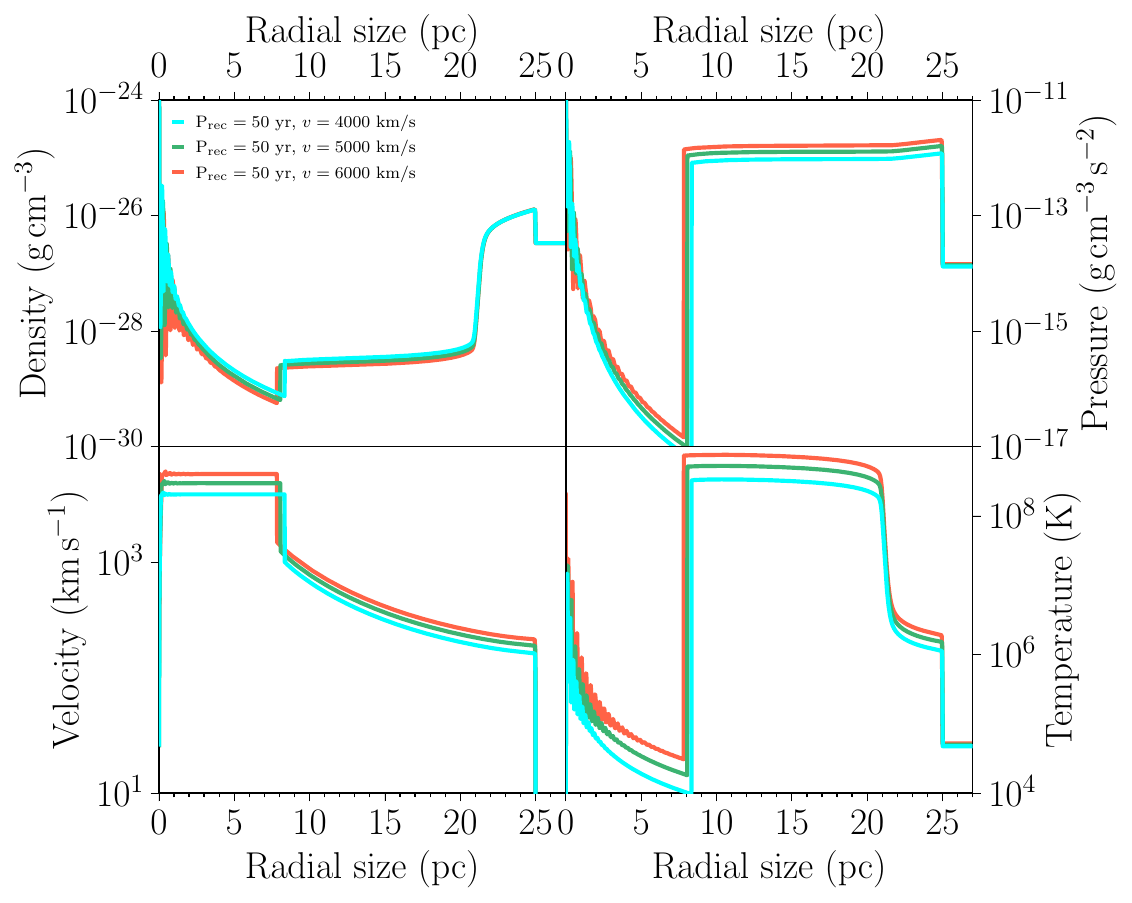}
\caption{Comparing the dynamics of the NSRs grown in Runs 1 -- 15. Top left: Dynamics of remnants grown in Runs 1 -- 5 (including the reference simulation, Run 4) with a constant $v_{\text{ej}} = 6000 \ \text{km} \ \text{s}^{-1}$ and varying $\text{P}_{\text{rec}}$. Top right: Dynamics of remnants grown in Runs 6 -- 10 with a constant $v_{\text{ej}} = 5000 \ \text{km} \ \text{s}^{-1}$ and varying $\text{P}_{\text{rec}}$. Bottom left: Dynamics of remnants grown in Runs 11 -- 15 with a constant $v_{\text{ej}} = 4000 \ \text{km} \ \text{s}^{-1}$ and varying $\text{P}_{\text{rec}}$. Bottom right: Dynamics of remnants grown in Runs 4, 9 and 14 with a constant $\text{P}_{\text{rec}} = 50$ years and varying ejecta velocity.}
\label{dynamics all runs}
\end{figure*}
\subsection{Bulk ejecta: Runs 1 -- 3, 5 -- 15}\label{sec:Other bulk runs}
After establishing the dynamics of a reference simulation (Run 4), we explored NSR evolution from a KT\,Eri-like system with varying recurrence periods and ejecta velocities: Runs 1--3 and Runs 5--15. The parameters used in these runs are provided in Table~\ref{Runs} and we show the results of Runs 1 -- 15 (with the inclusion of the reference simulation) in Figure~\ref{dynamics all runs}.

It is immediately apparent from the top left ($6000 \ \text{km} \ \text{s}^{-1}$ in Runs 1 -- 5), top right ($5000 \ \text{km} \ \text{s}^{-1}$ in Runs 6 -- 10) and bottom left ($4000 \ \text{km} \ \text{s}^{-1}$ in Runs 11 -- 15) plot of Figure~\ref{dynamics all runs} that varying the recurrence period does not affect the large-scale structure of the NSR. Specifically, the cavity/pile-up region boundary and the inner and outer edge of the shell extend out to the same radius in all runs with the same ejecta velocity. However, the longer the period between eruptions, the more structure we see in the cavity (along with a marginal amount of fluctuation in the inner ejecta pile-up region) as a result of resolving the individual eruptions. Within the cavity, the average density and velocity remain the same as the recurrence period is altered, yet the pressure and therefore temperature within the region increase moderately as $\text{P}_{\text{rec}}$ becomes longer. Furthermore, the higher velocity ejecta drive larger fluctuations in density, pressure and temperature, likely because of the higher levels of KE being ejected from each nova eruption.

The bottom right plot of Figure~\ref{dynamics all runs} illustrates the fully grown NSR from Runs 4, 9 and 14: systems with the same recurrence period of 50 years but with ejecta velocities of $6000 \ \text{km} \ \text{s}^{-1}$, $5000 \ \text{km} \ \text{s}^{-1}$ and $4000 \ \text{km} \ \text{s}^{-1}$, respectively. As previously found, the inner and outer edge of the NSR shells all extend to the same distance of ${\sim}21.6$ pc and 25 pc, respectively, corresponding to a thickness ${\sim}13.7$ per cent (as in the reference simulation in Section~\ref{sec:Reference simulation dynamics}). Though, in contrast to the similarity in the dynamics of the NSRs when sampling the recurrence period (with only the cavity showing minor differences), varying the ejecta velocity produces more noticeable changes (as shown in the bottom right-hand plot of Figure~\ref{dynamics all runs}).
\begin{figure*}
\centering
\includegraphics[width=0.49\textwidth]{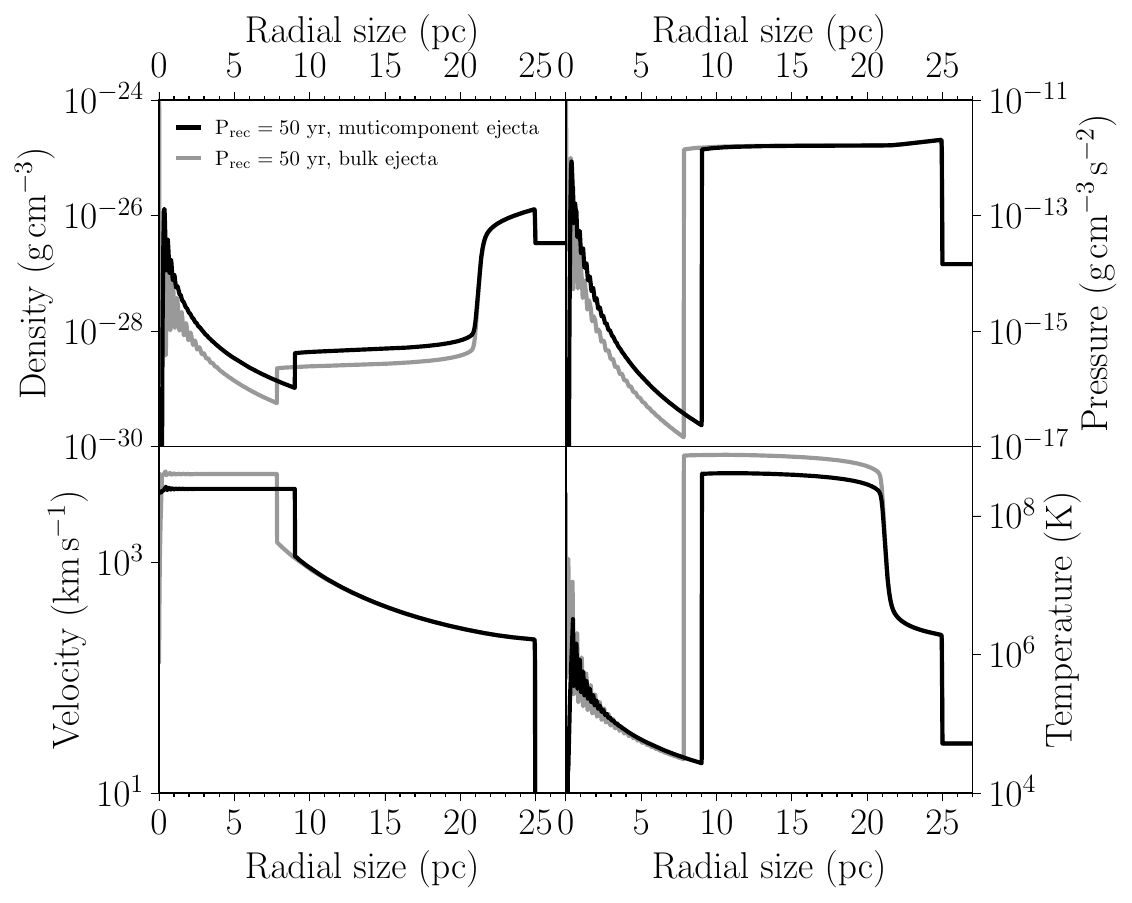} \quad
\centering
\includegraphics[width=0.49\textwidth]{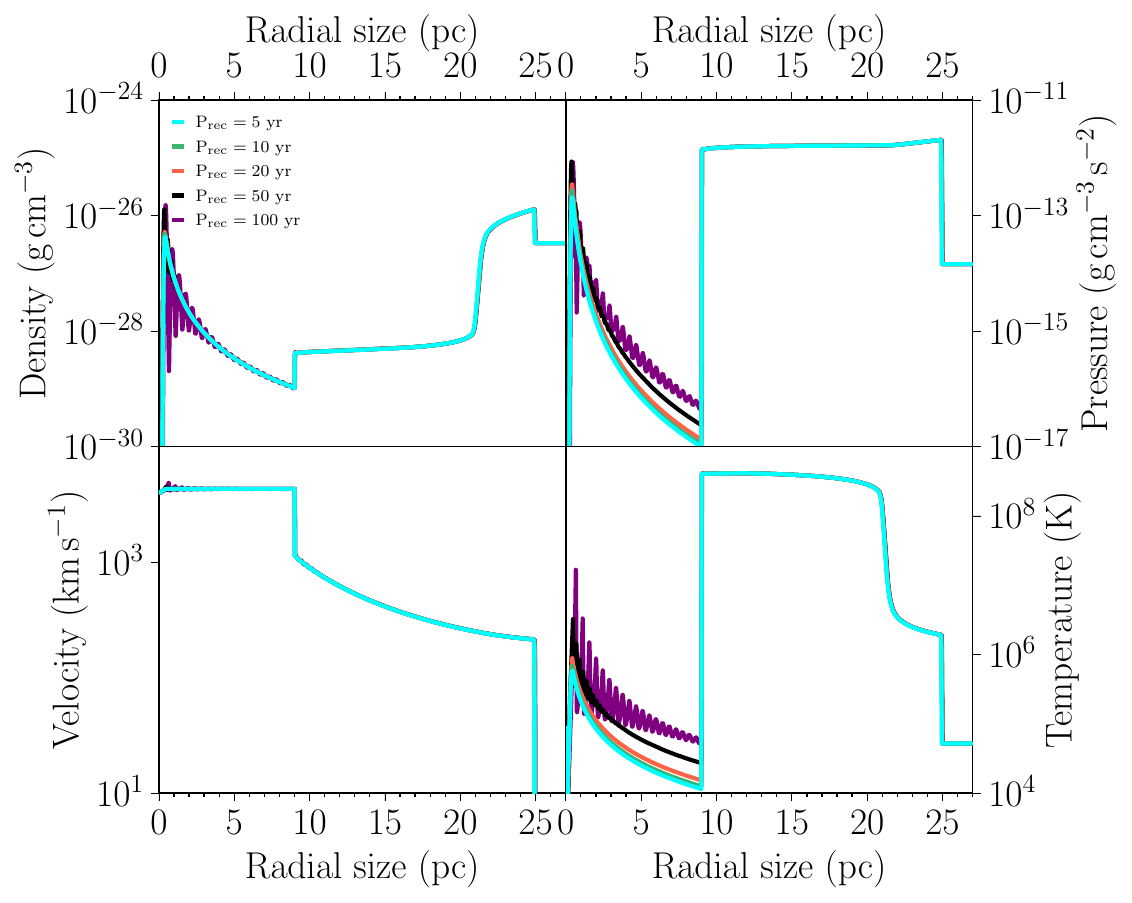}
\caption{Comparing the dynamics of the NSRs grown in Runs 1$^{\star}$ -- 5$^{\star}$. Left: Dynamics of the NSR grown in the Run 4 (reference simulation) compared to Run 4$^{\star}$. Run 4 is denoted as the model with bulk ejecta and Run 4$^{\star}$ is denoted as the model with multicomponent ejecta. Right: Dynamics of remnants grown with multicomponent ejecta with varying recurrence periods (1$^{\star}$ -- 5$^{\star}$).}
\label{mcomp dynamics}
\end{figure*}

Specifically, the radial extent of the cavity/pile-up region boundary increases from ${\sim}7.8$ pc to ${\sim}8.1$ pc to ${\sim}8.3$ pc as the ejecta velocity decreases ($6000 \ \text{km} \ \text{s}^{-1}$ to $5000 \ \text{km} \ \text{s}^{-1}$ to $4000 \ \text{km} \ \text{s}^{-1}$) -- this likely results from the increasing number of eruptions (and therefore the longer evolutionary times) needed to reach the radius of 25 pc also allowing more time to move the cavity border further. As we would expect, the velocity of the ejecta traversing the cavity is lower for Run 14 (${\sim}3860 \ \text{km} \ \text{s}^{-1}$) than Run 4 (${\sim}5800 \ \text{km} \ \text{s}^{-1}$) on account of the lower initial velocity for the former -- the small drop in ejecta velocity seen in all runs is directly related to the small (but non-negligible) resistance of the lower density material in the forming cavity. The difference in the velocity profile for Runs 4, 9 and 14 is maintained throughout the pile-up region and across the NSR shell.

In the bottom right-hand panel of the bottom right-hand plot in Figure~\ref{dynamics all runs} we see that the temperature across the whole NSR is cooler for lower velocity ejecta. For example, the temperature of the inner ejecta pile-up (on the border with the cavity) for Run 14 (the cyan line) is ${\sim}3.2 \times 10^8$\,K whereas the same region in Run 4 (the red line) reaches ${\sim}7.1 \times 10^8$\,K. This large difference in temperature is also found in the NSR shell, with the lower velocity ejecta leading to cooler shells.

\subsection{Multi-component ejecta}\label{sec:Multicomponent runs}
Here, we consider the NSRs grown with ejecta composed of multiple components. As detailed in Table~\ref{Runs}, and by design (as the KE of each multi-component ejecta matches the KE of the bulk ejecta), Run 4$^{\star}$ takes the same time (${\sim}5.1 \times 10^4$ years) as Run 4 (both with $\text{P}_{\text{rec}} = 50$ yr) to grow a NSR to match the size of observed KT\,Eri shell. The nova in this model also exhibits an identical number of eruptions (1,019) to match the NSR in Run 4 and as a result ejects the amount of kinetic energy as the reference simulation.

We grew the NSR in Run 4$^{\star}$ until it reached the size of the observed shell presented in Shara et al.\ (2023) as we have with all other models in this study. Yet, despite reaching this same size (and having the same shell thickness) as Run 4, the NSR in Run 4$^{\star}$ displays appreciably different dynamics. As illustrated in the top left panel of the left-hand plot of Figure~\ref{mcomp dynamics}, the boundary between the cavity and the ejecta pile-up region is situated ${\sim}1.2$ parsecs further from the central nova compared to the equivalent border in Run 4. Furthermore, the density within the cavity and pile-up region is approximately one order of magnitude higher in the NSR grown with multicomponent ejecta compared to the NSR grown with bulk ejecta. While the pressure and velocity are identical across the ejecta pile-up and NSR shell, the pressure in the cavity is higher ($\times2.5$) in Run 4$^{\star}$ compared to Run 4 and the velocity is lower ($\times0.75$). As shown in the bottom right panel of the left-hand plot in Figure~\ref{mcomp dynamics}, the temperature in the cavity and shell are very similar whereas there is a large drop in the temperature of the whole pile-up region. The differences in the dynamics outlined above arise from the lower density components of the ejecta in Run 4$^{\star}$ being more susceptible to radiative cooling and also the slower gas, which takes longer to traverse the cavity.

In the right-hand plot of Figure~\ref{mcomp dynamics}, we show the dynamics of the remnants grown with multicomponent ejecta with varying recurrence periods in (Runs 1$^{\star}$ -- 5$^{\star}$; see Table~\ref{Runs} for details). In a similar manner to the NSRs grown with bulk ejecta eruptions of the same ejecta velocity, we see that the remnants in Runs 1$^{\star}$ -- 5$^{\star}$ all share the same (almost identical) structure beyond the ejecta pile-up boundary. While the velocity remains the same in the cavity for all multicomponent ejecta runs, the pressure and temperature in this region do vary between runs. Additionally, the density within the cavity regions are largely different: the shorter the recurrence period, the smoother the density distribution. Large fluctuations apparent in the NSR cavity grown from a nova with a long recurrence period reveal individual nova ejecta traversing this region before their collision with the ejecta pile-up boundary.

\begin{figure}
\includegraphics[width=\columnwidth]{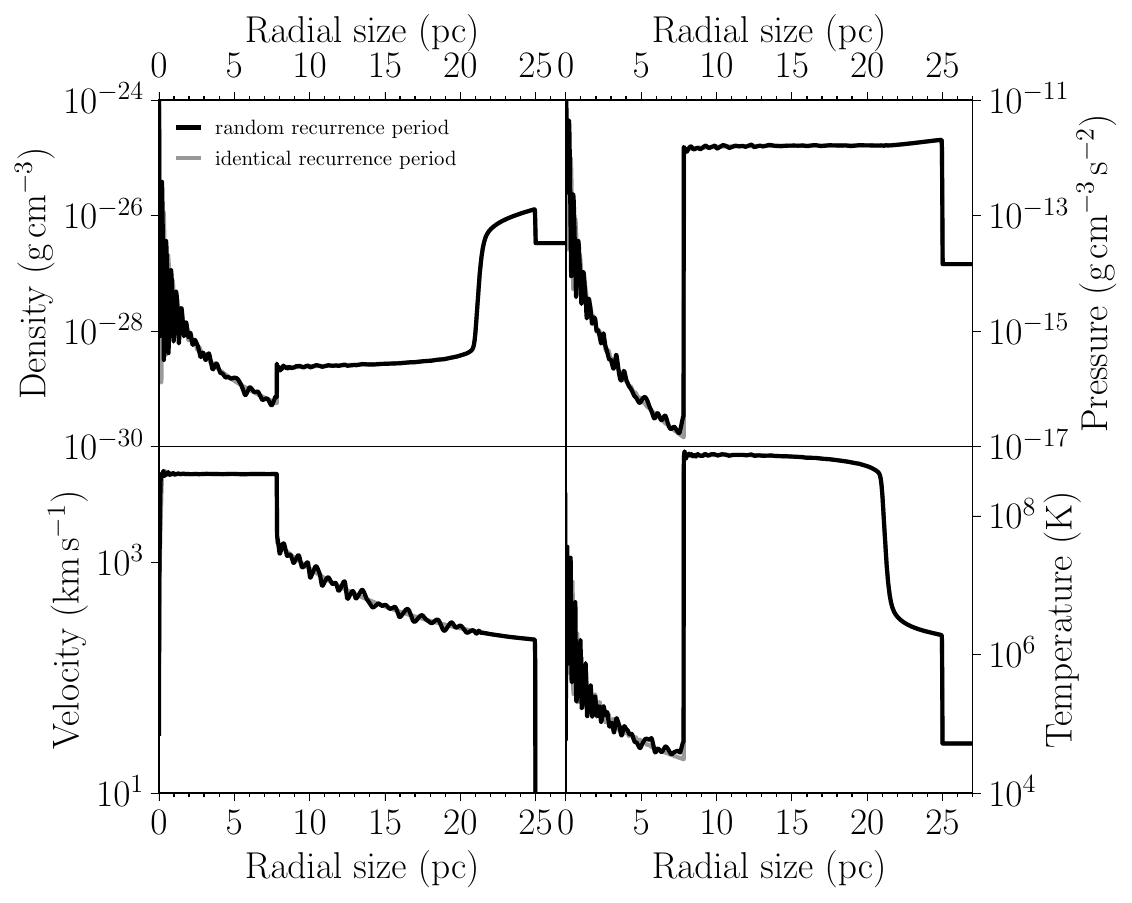}
\caption{Comparing the dynamics of the NSR from Run 4 (reference simulation with smoothly distributed eruptions) with the dynamics of the NSR from Run 4$^{\dag}$ with Gaussian distributed eruptions. \label{fig:compare ref with random}}
\end{figure}
\subsection{Randomly occurring eruptions}\label{sec:Gaussian distribution recurrence periods}
For Runs 1 -- 15 and Runs 1$^{\star}$ -- 5$^{\star}$, we have a recurrent nova evolving with constant recurrence period. However, known RNe deviate from their average recurrence period by up to 10 per cent \citep{2018ApJ...857...68H,2021gacv.workE..44D}. To implement this feature, we allowed each inter-eruption period to differ by randomly selecting the inter-eruption times from a Gaussian distribution with a mean recurrence period of 50 years ($\mu = 50$) and a standard deviation of 10 per cent ($\sigma = 0.1$) and constructed the corresponding nova ejecta.

As shown in Figure~\ref{fig:compare ref with random}, random inter-eruption periods selected from a Gaussian distribution ($\mu = 50$, $\sigma = 0.1$) between eruptions has no significant impact on the structure of the NSR (Run 4$^{\dag}$) compared to the reference simulation (Run 4; with identical inter-eruptions times). The number of eruptions (1,033) and total time of evolution ($5.2 \times 10^4$ years) closely match the values for the reference simulation (see Table~\ref{Runs}), as does the total kinetic energy. As the same amount of energy is being injected from the central nova into the surrounding ISM, it is inevitable that a NSR with a radius of 25 pc is created in a similar time frame.

The one difference between the two NSRs is the development of some structure in the cavity and inner ejecta-pile up region. There are small fluctuations in density, pressure and temperature in the cavity region whereas the more significant fluctuations in velocity are located within the ejecta-pile up region. As the material is being ejected from the nova at non-identical time intervals in this scenario, a proportion of the ejecta will catch the preceding ejecta at random points of the NSR - this gives rise to the non-smooth distribution of material within the structure. However, as with all the other simulations, the structure of the NSR shell is unaffected.

\section{Discussion}\label{sec:Discussion}
For the recurrence periods sampled in our study, we find that the NSR shell grown from ejecta with the same velocity reach a radius of 25 parsecs in the same amount of time, with only the number of eruptions changing. As such, the time taken for the NSR to grow to the observed size of the shell found surrounding KT\,Eri (Shara et al.\ 2023) in our reference simulation (${\sim}50$\,kyr) is fixed regardless of the period between eruptions. A 50 kyr timescale also supports the assumption that the WD in the KT\,Eri system is relatively young compared to other RNe.

Likewise, the shell thickness of ${\sim}14$ per cent for the reference simulation NSR is consistent across all runs employing eruptions with bulk ejecta. Though the ISM surrounding KT\,Eri is likely to be non-uniform and so shapes the observed NSR into a complex structure, the predicted thickness of ${\sim}14$ per cent from our model is close to the thickness of northern edge of the observed shell (Shara et al.\ 2023). Though, unlike in \citet{2023MNRAS.521.3004H}, we have not considered mass-loss occurring during prior phases of KT Eri's evolution. The compression of the NSR shell in the direction of proper motion (see Fig 3 of Shara et al.\ 2023) may then be evidence of interaction with a bow shock from this earlier phase. Also, magnetic fields, not considered in this work, affect stellar winds which then have a large impact on ISM structure \citep{2016MNRAS.459.1803W} and consequently the growth of a NSR shell.

In Section~\ref{sec:Multicomponent runs}, we investigated the impact of velocity variations within the nova ejecta on the structure of the growing NSR. Even though we tuned the model such that the final NSR shell extended to 25 parsecs, we did find that the NSR substructure is considerably different compared to the NSR grown with constant ejecta velocity eruptions. This is a consequence of the components with slower, low-density gas being initially cooler and therefore losing energy through radiative cooling more efficiently than the bulk ejecta in the reference simulation (Run 4). In \citet{2023MNRAS.521.3004H}, we demonstrated that a `two-component' (rather than multiple component as tested here) ejecta did not have an discernible effect on the NSR structure as, in this scenario, the ejecta still met the criteria whereby they were not able to cool efficiently. The twelve-component ejecta (in Runs 1$^{\star}$ -- 5$^{\star}$) are more complex and therefore we see a more dramatic difference in the NSR sub-structure due to more efficient cooling.

\section{Conclusions}\label{sec:Conclusions}
We have presented a set of simulations used to model a nova super-remnant found in narrow-band imaging surrounding the Galactic classical nova, KT\,Eridani (Shara et al.\ 2023). For the models with $v_{\text{ej}} = 6000 \ \text{km} \ \text{s}^{-1}$, a nova super-remnant was grown to match the radial size of the observed structure (${\sim}25$ pc) in ${\sim}$50,000 years -- consistent with the proper motion history of KT\,Eri (Shara et al.\ 2023). Additionally, the NSR shell thickness of ${\sim}14$ per cent found in the models is in reasonable agreement with the observed shell.

An estimate of the integrated X-ray luminosity of the KT\,Eri nova super-remnant suggests that the structure may be accessible to wide-field X-ray facilities. Our prediction of the H$\alpha$ surface brightness of the NSR shell is ${\sim}20-50$ times larger than that measured by Shara et al.\ (2023), however, this will, at least, in part be due to several of our assumptions, most importantly that of spherical symmetry. Exploring cooling conditions and implementing complex ISM structure through differing mass-loss phases and magnetic field shaping in future work will help to reconcile our model predictions with observed parameters.

We strongly encourage further observations of the KT\,Eri nova super-remnant, especially exploration of the predicted X-ray emission. 

\section*{Acknowledgements}
The authors would like to thank our reviewer, Dr Christopher Wareing, for his helpful suggestions that we will implement in future modelling. MWH-K acknowledges a PDRA position funded by the UK Science and Technology Facilities Council (STFC). MWH-K and MJD receive funding from STFC grant number ST/S505559/1. MMS, KML and JTG acknowledge the support of NSF award 2108234. This work made use of the high performance computing facilities at Liverpool John Moores University, partly funded by LJMU’s Faculty of Engineering and Technology and by the Royal Society.

\section*{Data Availability}
The data in this study can be shared on reasonable request to the corresponding author. This work was conducted with the \texttt{Morpheus} \citep{2007ApJ...665..654V} program and analysed using the Python libraries: Numpy \citep{harris2020array} and Matplotlib \citep{Hunter:2007}.

\bibliographystyle{mnras}
\bibliography{bibliography}
\balance 

\bsp
\label{lastpage}
\end{document}